\documentclass[sigconf,authorversion,nonacm]{acmart}

\usepackage{graphicx}
\usepackage{balance}
\usepackage{hyperref}
\usepackage{booktabs}
\usepackage{hyperref}
\usepackage{verbatim}
\usepackage{tikz}
\usepackage{graphicx}
\usepackage{cellspace}
\usepackage[htt]{hyphenat}
\usepackage{lipsum,graphicx,subcaption}
\usepackage{multicol}
\usepackage{multirow}
\usepackage{paralist}
\usepackage{breakcites}
\usepackage{amsmath,comment}
\usepackage[autostyle]{csquotes}
\usepackage{pbox}
\usepackage{booktabs}
\usepackage{enumitem}
\usepackage{booktabs}
\usepackage{pifont}
\graphicspath{{./figures/}}

\usepackage{amssymb}

\newcommand{\bpar}[1]{\smallskip\noindent\textbf{{#1}}}

\usepackage{url}
\usepackage{fancyhdr}
\fancypagestyle{firstpage}{%
  \lhead{}
  \chead{Paper accepted for publication at AsiaCCS 2021}
}

\begin{document}
\title{MalPhase: Fine-Grained Malware Detection Using Network Flow Data}

\author{Michal Piskozub}
\affiliation{%
  \institution{University of Oxford}
  \city{Oxford}
  \country{UK}
}
\email{michal.piskozub@cs.ox.ac.uk}
\par
\author{Fabio De Gaspari}
\affiliation{%
  \institution{Sapienza University of Rome}
  \city{Rome}
  \country{Italy}
}
\email{degaspari@di.uniroma1.it}

\author{Frederick Barr-Smith}
\affiliation{%
  \institution{University of Oxford}
  \city{Oxford}
  \country{UK}
}
\email{freddie.barr-smith@cs.ox.ac.uk}

\author{Luigi V. Mancini}
\affiliation{%
  \institution{Sapienza University of Rome}
  \city{Rome}
  \country{Italy}
}
\email{mancini@di.uniroma1.it}

\author{Ivan Martinovic}
\affiliation{%
  \institution{University of Oxford}
  \city{Oxford}
  \country{UK}
}
\email{ivan.martinovic@cs.ox.ac.uk}

\begin{abstract}
Economic incentives encourage malware authors to constantly develop new, increasingly complex malware to steal sensitive data or blackmail individuals and companies into paying large ransoms. In 2017, the worldwide economic impact of cyberattacks is estimated to be between 445 and 600 billion USD, or $0.8\%$ of global GDP\footnote{Report from CSIS, McAfee: \url{https://www.mcafee.com/enterprise/en-us/assets/reports/restricted/rp-economic-impact-cybercrime.pdf}}. 
Traditionally, one of the approaches used to defend against malware is network traffic analysis, which relies on network data to detect the presence of potentially malicious software. However, to keep up with increasing network speeds and amount of traffic, network analysis is generally limited to work on aggregated network data, which is traditionally challenging and yields mixed results. In this paper we present MalPhase, a system that was designed to cope with the limitations of aggregated flows. MalPhase features a multi-phase pipeline for malware detection, type and family classification. The use of an extended set of network flow features and a simultaneous multi-tier architecture facilitates a performance improvement for deep learning models, making them able to detect malicious flows ($>98\%$ F1) and categorize them to a respective malware type ($>93\%$ F1) and family ($>91\%$ F1). Furthermore, the use of robust features and denoising autoencoders allows MalPhase to perform well on samples with varying amounts of benign traffic mixed in. Finally, MalPhase detects unseen malware samples with performance comparable to that of known samples, even when interlaced with benign flows to reflect realistic network environments.
\end{abstract}

\maketitle

\thispagestyle{firstpage}

\section{Introduction}
The quantity of new, unique malware samples has greatly increased over the years, from approximately six new daily samples in 2000~\cite{10.5555/517959} to what a recent study estimates at 344,041 unique daily samples in 2018~\cite{Ugarte-Pedrero2018}. Today, new malware families still find widespread success, with malware campaigns providing large economic incentives to their creators and recent attacks such as WannaCry and Zeus resulting in billions of dollars of losses. Many techniques have been proposed in response to this rapid growth in daily threats, ranging from traditional signature-based detection methods, to more recent approaches based on machine learning and behavioural modeling (for a good survey on the subject, the reader is referred to~\cite{7307098}). 
The sub-field of malware network traffic analysis focuses on methods to detect malware using only network-level information. Network-based approaches tend to provide simpler scalability and maintainability compared to holistic behavioural analysis approaches, as network traffic can be easily captured and analysed at a single (or a few) points in the network. 
Network-level analysis is also more resilient to malware that attempts to identify and evade the detection system itself, as analysis is usually performed by Network Intrusion Detection Systems (NIDS) that are physically separated from the monitored systems.

Over the last few decades, researchers have proposed many different network traffic analysis approaches which can identify ongoing malware communication in the network, with varying degrees of success. Based on the level of granularity of the information they require, these approaches can generally be categorized as \emph{packet-level} network analysis and \emph{flow-level} network analysis. Packet-level network analysis approaches use detailed network traffic data such as individual packets or HTTP connections in order to classify network streams~\cite{Ahmed2011}.
Packet-level approaches can generally reach very high performance, as the use of detailed network traces enables the detection of finer differences between benign and malware traffic, that are otherwise lost when using aggregated network data (e.g., NetFlows~\cite{Alahmadi2020,Piskozub2019}). However, the use of detailed network traces renders these approaches ill-suited to real life applications, where capturing such traces at line speed is challenging and expensive. Flow-level network analysis approaches on the other hand use aggregated traffic flow information, which can be more readily logged for analysis. However, classification based on flows is considerably more difficult, with traditionally mixed results due to the reduced amount of information available. Approaches that rely on network flows generally have reduced scope and are limited to malware detection only~\cite{Bilge2012}, malware type classification with few families~\cite{Piskozub2019} or family classification only for a specific malware type~\cite{Alahmadi2020}.

\begin{figure*}[t]
    \centering
    \includegraphics[width=\textwidth]{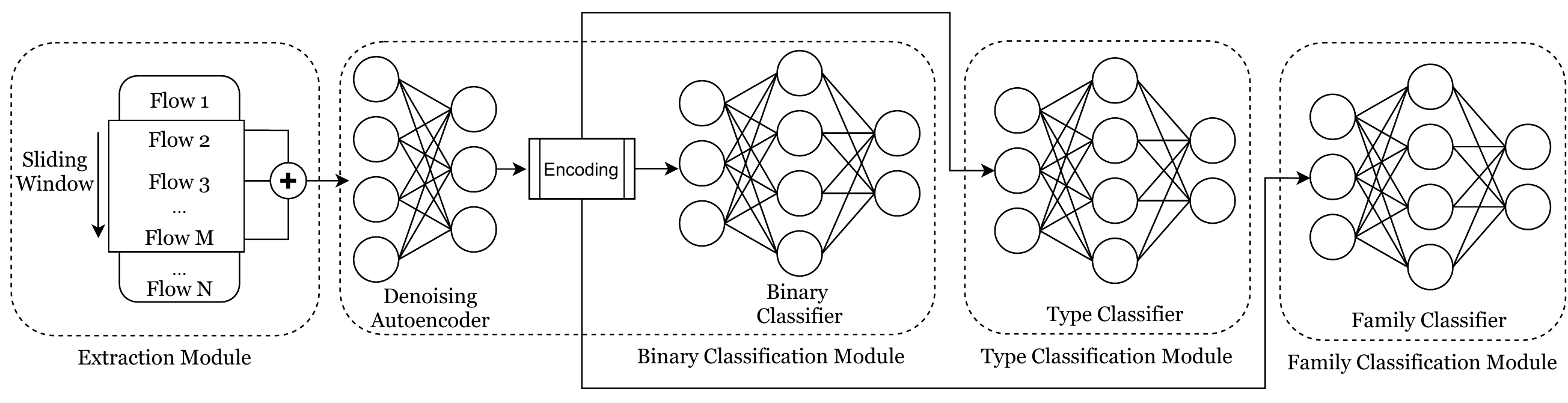}
    \caption{Detailed architecture of MalPhase. A sliding window extracts $M$ flows, which are encoded and compressed in a compact representation. The encoded flows are ran through a denoising autoencoder to filter-out benign traffic noise. The remaining malware features are passed on to the binary classifier.  and successively to the type and family classifiers if the sample is malicious.}
    \label{fig:malphase_architecture_detail}
\end{figure*}

Using one of the largest datasets of malware traffic to date, in this paper we revisit the topic of traffic analysis with aggregated flows. We believe that the large amount of malware network data that is available today from sources such as VirusTotal, combined with the ability of neural networks to extract robust features from such a large dataset, can allow to overcome the limitations of flow-based analysis.
We propose MalPhase, a multi-phase system that is designed to cope with the limitations of flow-based analysis and is able to detect malware traffic, as well as classify it to a specific malware type and malware family. MalPhase uses a multi-tier classification system that is responsive to short term bursts of malware traffic, as well as slower, more regular malware network activity. 
Our evaluation demonstrates that MalPhase achieves extremely high performance in malware detection that is comparable or better than state-of-the-art~\cite{Piskozub2019,Alahmadi2020} ($>98\%$ F1-score), while also broadening the scope of classification to malware type ($>93\%$ F1-score) and malware family ($>91\%$ aggregated F1-score). Finally, MalPhase uses a combination of denoising autoencoders and Deep Neural Network (DNN) classifiers that make it resilient to high levels of noise injected in the malware traffic, as well as being able to detect unseen malware samples with performance comparable to that of known samples.

\smallskip\noindent
\textbf{Contributions.} 
In this paper we make the following core contributions:
\begin{enumerate}
    \item We design and implement MalPhase, a multi-phase system for malware detection and classification based on network flows. MalPhase is able to detect a large set of malware families from network flows, as well as classify it to a malware type and even specific malware family. 
    \item The MalPhase system can work both as an offline analysis tool, as well as online on live data. This is enabled by its multi-tier design which allows it to work on different timescales in parallel.
	\item We evaluate MalPhase on a large, standardized malware dataset, obtained from a single source. To the best of our knowledge, this is one of the largest datasets of labeled malware traffic to date ($\sim 1$ billion flows). We assess MalPhase performance in a real world setting composed of both clean malware traffic as well as noisy traffic (i.e., malware traffic with mixed in benign flows), 
	assessing the robustness of our system to real life conditions.  We also evaluate the ability of MalPhase to generalise and detect unseen malware samples, even with the presence of noise.
\end{enumerate}

\section{System Architecture}
\label{sec:system_architecture}

MalPhase is a multi-tier, multi-phase system that combines supervised and unsupervised machine learning techniques to provide fine-grained malware classification from aggregated network flow data. Rather than analysing traffic network-wide, MalPhase works on a per-host basis and supports both real-time traffic analysis as well as analysis of stored network data.
As depicted in Figure~\ref{fig:malphase_architecture_detail}, the MalPhase pipeline is comprised of a flow extraction and encoding step, followed by three classification phases:

\begin{enumerate}
    \item \textbf{Binary Classification Phase.} In the first phase, the network flows undergo a \emph{denoising} step aimed at extracting features linked to malicious activity, while filtering out features arising from benign traffic. The denoised flows are then input to a binary DNN classifier which categorizes them as either \emph{benign} or \emph{malicious}.
    
	\item \textbf{Malware Type Classification Phase.} Flows categorized as malicious in the binary phase are further input to a multiclass DNN classifier which identifies the \emph{type} of malware that generated the flows (e.g., worm, ransomware). Consistent with other categorisations in the literature~\cite{ciscoMalware,Piskozub2019}, in MalPhase we consider five types of malware: adware, ransomware, trojan, virus, and worm.
	
	\item \textbf{Malware Family Classification Phase.} In the third and final phase, malicious flows are fed to a type-specific DNN classifier (e.g., ransomware classifier if the flows are classified as belonging to a ransomware in phase 2), which further categorizes them into the specific malware family that generated the traffic.
\end{enumerate}

We designed the architecture of MalPhase classification modules  to provide robust classification that is resilient to noisy inputs, due to the use of denoising autoencoders, as well as to different distributions of benign and malicious traffic, thanks to the use of a multi-tier system. In the following sections, we discuss in detail the different components of MalPhase.

\subsection{System Overview}
\label{sec:system_overview}

The MalPhase pipeline is comprised of two main modules: (1) the extraction module and (2) the classification modules. The extraction module is responsible for extracting flow features from a window of $m$ flows, encoding and concatenating them before forwarding the compacted features to the classification module. The classification modules are responsible for denoising the encoded flows, removing features related to benign traffic, and for classifying the flows with the appropriate classifier based on the current phase (binary, type or family). A detailed illustration of this process is depicted in Figure~\ref{fig:malphase_architecture_detail}. As previously discussed, MalPhase's pipeline uses three classification modules, each handling a different phase of classification: (I) benign/malicious classification; (II) malware type classification; (III) malware family classification. The MalPhase pipeline is designed to work on a per-host level, rather than at a network-wide level. This allows MalPhase's classification modules to deal with much less noise in the form of benign traffic generated by other machines, which would make it much harder to detect and classify malware traffic.

The MalPhase pipeline is replicated and structured in an $N$-tier system: each tier runs a full copy of the pipeline on a different group of flows: lower tiers work on short sequences of flows, while higher tiers process increasingly longer sequences of flows.
The classification modules of each tier are fine-tuned to work on a specific number of flows and require different amounts of malware flows to accurately classify samples, with higher tiers requiring more flows than lower tiers. We design this tier system to allow MalPhase to monitor both short-term and long-term behavioural changes in flows: lower tiers capture short bursts of malicious flows, while higher tiers capture slower, more regularly-occurring malicious flows, similarly to~\cite{continella2016shieldfs}. Moreover, the tiered architecture of MalPhase supports both real-time traffic analysis and analysis of stored flow data. Since lower tier classification modules operate on shorter flow windows, they provide faster response times to malware activity compared to higher tiers. On the other hand, higher tier classification modules are much more resilient to noise and are more precise than lower tier modules with non-burst malicious flows, as we will see in Section~\ref{sec:mixed_eval}.

\bpar{Flow Extraction and Encoding}\\
The extraction module of MalPhase uses multiple sliding windows, one for each tier, to extract a subset of flows from a larger sequence. Each individual window is then encoded before being forwarded to the tier's classification module. The encoding step filters the raw network flows, excluding or transforming features that could bias the classifier. For instance, most malware are coded to communicate with a fixed set of IP addresses to download additional malicious code, or to exfiltrate stolen data to the malware author. If a classifier was trained with IP address as a feature, it is likely that the model would assign a heavy weight to it and become biased, since IPs tend to be unique per malware family. While such a classifier would probably exhibit high performance on the testing portion of the datset it was trained on, it wouldn't be able to maintain such performance if a new version of the same malware family were to use different IPs for communication.
For these reasons, in the encoding phase we only use features that are harder to spoof and linked to the behaviour of the malware itself. We use features that are included in the standard definition of a network flow, extended by the round-trip time and Shannon entropy of packet payloads. In particular, we extract the following flow features:
flow duration, round-trip time, IP protocol used, connection towards local or public IP, destination port, packets sent, bytes sent, packets received, bytes received, sent packet payload entropy, and received packet payload entropy.

\begin{table*}[t]
\small
\centering
\begin{tabular}{l l r r r c}
\toprule
\textbf{Access} & \textbf{Dataset} & \textbf{Hashes} & \textbf{VirusTotal Pcaps} & \textbf {AVClass Pcaps} & \textbf{Flows}   \\
\midrule
public & ClamAV~\cite{clamav} & 3,447,223 & 1,289,442 & 1,214,760 & 229,045,355\\
public & Ember Malicious~\cite{anderson2018ember} & 800,000 & 453,467 & 436,166 & 26,568,413\\
quasi-restricted & GT Malware~\cite{gt} & 12,739,759 & 10,288,165 & 10,080,539 & 658,835,290\\
public & MalRec~\cite{malrec} & 66,292 & 40,731 & 37,776 & 12,273,572\\
public & MalShare~\cite{malshare} & 3,410,439 & 1,313,244 & 1,274,333 & 32,311,660\\
mixed & Misc. & 1,955,923 & 937,690 & 648,570 & 28,784,436\\
public & VirusShare~\cite{virusshare} & 3,147,748 & 2,339,362 & 2,182,300 & 43,810,096\\
public & VX Underground~\cite{vxunderground} & 405,803 & 95,023 & 87,629 & 1,489,459\\
\midrule
\multicolumn{2}{c}{Total (unique)} & 13,920,730 & 15,369,738 & 14,644,135 & 996,712,027\\
\bottomrule
\end{tabular}
\caption{Overview of malicious datasets.}
\label{malwareDatasets}
\vspace{-5mm}
\end{table*}

\subsection{Denoising and Classification}
\label{sec:denoising}
In a real world setting, the traffic generated by an infected host will include both malicious flows and flows generated by benign programs at the same time. MalPhase is designed to be resilient to varying amounts of benign flows (called \emph{noise}) in the network traffic, thanks to the use of a denoising autoencoder~\cite{vincent2010stacked}. Denoising autoencoders are a type of autoencoder that is trained to remove noise from samples belonging to a given distribution. Denoising autoencoders are notably used in image classification, where they are able to filter surprising amounts of noise and reconstruct the underlying image~\cite{7836672,NIPS2012_4686}. Similarly, in MalPhase each tier uses a denoising autoencoder to filter benign noise and extract clean malware features, which are then passed to the DNN classifier of the appropriate phase. It is worth noting that the autoencoder does not simply remove benign flows from a given sample, leaving only malware flows. Rather, it compresses a sample in a robust, compact representation that preserves malware-related flow features, while filtering out most benign-related flow features. It is this compact representation that is then used by the DNN classifiers. In MalPhase, the autoencoders are trained by taking as input noisy malicious traffic samples, obtained by randomly injecting benign flows in a purely malicious sample, and learn to output the original clean sample without noise. At inference time, only the encoder part of the autoencoder is used, which is what provides the compact feature representation discussed above (see Figure~\ref{fig:malphase_architecture_detail}).
\section{Datasets}
\label{sec:dataset}

\begin{table}
\small
\centering
\begin{tabular}{l l r}
\toprule
\textbf{Access} & \textbf{Dataset} & \textbf{Flows} \\
\midrule
public & CTU~\cite{ctu} & 546,490\\
private & Research Server A & 70,319,185\\
private & Research Server B & 5,938,924\\
\midrule
\multicolumn{2}{c}{Total} & 76,804,599\\
\bottomrule
\end{tabular}
\caption{Overview of benign datasets.}
\label{benignDatasets}
\vspace{-10mm}
\end{table}
\begin{table*}[t]
\small
\centering
\begin{tabular}{l r r l}
\toprule
\textbf{Type} & \textbf{Hashes} & \textbf{Flows} & \textbf{Families (hashes \%, flows \%)}  \\
\midrule
adware & 1,117,311 & 26,198,428 & directdownloader (0.31\%, 7.12\%), downloadguide (24.96\%, 25.14\%),\\ 
 & & & hotbar (15.74\%, 11.4\%), inbox (7.41\%, 3.96\%), installcore (13.63\%, 17.42\%),\\ 
 & & & playtech (4.88\%, 6.17\%), softcnapp (30.85\%, 23.77\%), softonic (1.75\%, 1.72\%),\\
\vspace{0.75em}
 & & & techsnab (0.43\%, 3.26\%)\\
ransomware & 1,459,889 & 357,716,434 & cerber (0.89\%, 12.03\%), deshacop (0.01\%, 0.09\%), sage (0.01\%, 0.21\%),\\
\vspace{0.75em}
 & & & virlock (98.73\%, 87.41\%), wannacry (0.35\%, 0.23\%)\\
trojan & 2,063,369 & 276,427,311 & bublik (0.87\%, 2.15\%), byfh (0.31\%, 0.36\%), cycbot (0.11\%, 0.52\%),\\
 & & & delf (16.58\%, 10.66\%), mudrop (0.92\%, 0.97\%), ramnit (2.79\%, 0.3\%),\\
 & & & razy (2.6\%, 0.78\%), scar (3.7\%, 0.55\%), shiz (2.57\%, 9.36\%), ulise (3.5\%, 0.77\%),\\
 & & & unruy (18.19\%, 2.21\%), upatre (32.1\%, 66.62\%), vtflooder (6.08\%, 2.75\%),\\
\vspace{0.75em}
 & & & zbot (7.73\%, 1.42\%), zusy (1.88\%, 0.5\%)\\
\vspace{0.75em}
virus & 192,924 & 5,745,615 & pioneer (19.15\%, 15.29\%), sality (71.67\%, 76.32\%), viking (9.17\%, 8.37\%)\\
worm & 351,987 & 208,691,955 & allaple (22.51\%, 93.56\%), drolnux (4.77\%, 0.17\%), mydoom (23.91\%, 4.63\%),\\
 & & & socks (30.99\%, 0.93\%), warezov (8.67\%, 0.45\%), windef (9.12\%, 0.23\%)\\
\midrule
Total & 5,185,480 & 874,779,743 & \\
\bottomrule
\end{tabular}
\caption{Overview of families selected for the evaluation.}
\label{finalFamilies}
\vspace{-5mm}
\end{table*}
We make use of various datasets to train, test and evaluate our system. We acquire these datasets from public and private sources and seperate them into malicious and benign. In contrast to benign datasets, malicious datasets (Table~\ref{malwareDatasets}) contain hashes of malware binaries (from here on, \textit{hashes}). 

ClamAV is an open-source antivirus (AV) engine, which features a database of malware hashes to check against. We harvest most of its released hash lists to form the ClamAV dataset.
Ember is an open dataset of hashes of malicious Windows executables created as a benchmark for machine learning models. It contains malicious, benign and unlabeled hashes. For our needs, only malicious hashes are taken as Ember Malicious dataset.
The GT Malware dataset is a daily feed of flows that come from a number of newly observed malware by the Georgia Tech Information Security Center. The period considered in this paper ranges from 01.09.2018 to 30.11.2019. From this dataset we only use the published malware hashes, for which we download pcaps from VirusTotal, but do not use the associated network flows that are already part of this dataset. This is done for consistency reasons, as we use VirusTotal captures for all hash sources in order to work on a standardized dataset.
MalRec is a small dataset created as a result of a running malware on a dynamic analysis platform, which was collected over a two-year period.
MalShare is a community-driven open repository of malware samples, which features over 3 million hashes.
Similarly, VirusShare and VX Underground are virus sharing websites that aim to help security researchers analyze selected strains of malware.
The Miscellaneous (Misc.) dataset is created as an aggregation of hashes from a number of other public sources, too small to be considered as separate datasets.
As mentioned previously, we only collect malware hashes from the listed malware datasets.

The last row in Table~\ref{malwareDatasets} represents the sum of unique amounts of hashes, pcaps and flows. This is due to the fact that there are duplicate hashes across datasets, which then propagate to duplicated pcaps and flows. It is worth noting that the increase of total VirusTotal Pcaps with regards to total hashes is due to the fact that malware is often run in more than one sandbox and in those cases for one hash, there is more than one pcap. 
We check hashes of each malicious dataset against the VirusTotal database and download the corresponding AV reports, which list detection results from over 70 AV vendors, as well as network packet captures (pcaps) provided by their API. While these pcaps are a result of malware being detonated in a number of dynamic analysis sandboxes, not all of them are available on VirusTotal. This is due to the lack of binary files for a given hash uploaded by users to VirusTotal. Having this in mind, the number of VirusTotal pcaps in Table~\ref{malwareDatasets} is in some cases significantly smaller than the total number of hashes in a dataset. Following this trend, the number of usable pcaps is further reduced by feeding the VirusTotal AV reports to AVClass~\cite{10.1007/978-3-319-45719-2_11} - a malware labeling tool capable of computing malware family names from a number of labels given by AV engines. Since not all pcaps belong to a known family (to AV engines) or there is no clear consensus between AV labels, they are labeled as singletons and filtered out in the dataset creation process. 
As the last step, we convert pcaps to flows by using Yet Another Flowmeter (YAF)~\cite{yaf} - a suite of flow metering tools. The parameters used, ensure that a reasonable number of flows is obtained (--idle-timeout 30 --active-timeout 300), and additionally compute entropy of packet payloads (--max-payload 2048 --udp-payload --entropy). The resulting flows are in a bidirectional format, meaning that there are separate fields for forward and reverse directions of transferred number and sizes of packets, as well as their payload entropies.
What sets our described malicious dataset creation method apart is the fact that it comes from a single source, which makes it more standardized than the regular approach of combining different datasets captured under varying conditions and settings. By eliminating such variables, we improve the quality of the dataset, which translates to higher quality of trained models.

Benign network traffic is the second part of our dataset. As shown in Table~\ref{benignDatasets}, it comprises of a number of normal captures from the publicly-available CTU dataset from the Malware Capture Facility Project, and network traces from our research servers, spanning from 06.12.2019 to 18.07.2020 for server A and from 10.03.2020 to 18.07.2020 for server B. In the case of CTU, flows are created from pcaps, and in the case of our research servers we capture the flows directly, by using YAF with identical parameters to those outlined for the malware datasets. This ensures that all resulting flows in both malicious and benign datasets are homogeneous by being a product of the same tool used with matching parameters.

The Malicious flows, that we obtain by converting pcaps of known malware families (as identified by AVClass), are processed further by filtering out flows that are related to sandbox artefacts (i.e., traffic that was present in each packet capture, not related to the activity of malware). The resulting flows are assigned to family-based groups. We select those families that generate a sufficient number of flows for our classifiers. In the end, there are 38 malware families (Table~\ref{finalFamilies}), which we arrange into respective malware types by consulting reports from AV vendors. Additionally, we choose six malware families for the evaluation of MalPhase on unseen samples (Table~\ref{tab:ufamilies}). With regards to composition of the final malware dataset, we attempt to abide by the experimental best practices outlined by Rossow et al.~\cite{Rossow2012} concerning malware experiments.

\begin{table}
\small
\centering
\begin{tabular}{l r r}
\toprule
\textbf{Family} & \textbf{Hashes} & \textbf{Flows} \\
\midrule
autoit & 126,036 & 1,698,697 \\
banload & 17,141 & 609,371 \\
fareit & 49,445 & 709,474 \\
goldun & 496 & 390,957 \\
upantix & 11,368 & 511,520 \\
virut & 1,338,470 & 24,676,037 \\
\midrule
Total & 1,542,956 & 28,596,056 \\
\bottomrule
\end{tabular}
\caption{Overview of unseen families.}
\label{tab:ufamilies}
\vspace{-8mm}
\end{table}

\subsection{Noise Injection}
\label{sec:noisy_dataset}
In order to test the effectiveness of the denoising autoencoder approach and the resilience of the MalPhase architecture to noise, we created a noisy dataset by injecting benign flows in our original malware dataset. In the original, clean dataset, each malicious sample is comprised only of malware flows. In the noisy dataset we add to each malicious sample varying amounts of benign flows in random positions, simulating a more realistic network trace. 
In particular, given a malicious sample with $N$ malware flows, we add to it varying amounts of benign flows, increasing the size of the sample. We chose this method of injection for two reasons: 

\begin{compactitem}
\item It allows us to provide a more direct comparison between evaluation on clean samples and on mixed samples, since they both include the same amount of malware flows.
\item Adding benign flows on top of the existing malware flows is consistent with evasion techniques used by current malware~\cite{Tegeler2012a}. 
\end{compactitem}

The injected benign flows (from here on, referred to as \emph{noise}) are randomly sampled from a set of 76 million benign flows from our research servers and the CTU dataset. We use this noisy dataset in all experiments in Section~\ref{sec:mixed_eval}.

\section{Evaluation and Results}\label{eval}

In this section, we present our evaluation of MalPhase under three main sets of conditions, and aim at answering the following questions: 
(1) Given samples with either only benign or only malicious flows (\emph{clean samples}), what is the performance of each of MalPhase phases?
Evaluating MalPhase in ideal conditions allows us to define a baseline performance for the system. Whilst results on clean samples are not representative of real-world performance, they are indicative of the usefulness of MalPhase in a closed testing environment, such as a sandbox test for an unknown malware family, for instance. 
(2) How resilient is MalPhase to samples with mixed benign and malicious traffic (\emph{noisy samples})? 
In a real-world setting, a network traffic sample will contain both benign and malicious flows interleaved with each other. Evaluating the performance of MalPhase on noisy samples provides a better understanding of the system's performance in real life, where it is expected to be resilient to a high degree of noise generated by the various benign applications running beside the malware.  
(3) How well does MalPhase perform on unseen malware families? 
Given that new malware families emerge everyday, it is important that MalPhase can detect families that were not part of the training set, even when noise is injected. 
For all experiments in this section, each sample for a given MalPhase tier contains the following number of malicious flows: 

\begin{compactitem}
    \item Tier 1: $10$ malicious flows.
    \item Tier 2: $20$ malicious flows.
    \item Tier 3: $30$ malicious flows.
    \item Tier 4: $40$ malicious flows.
\end{compactitem}

\bpar{Training Details}\\
For all the experiments in this section, we trained the classification modules using the families presented in Table~\ref{finalFamilies}. In order to avoid excessive class imbalance, for families with a very large number of flows we only use a subset of the total flows, chosen by randomly sampling the overall set. The models used by the classifiers are feedforward neural networks using relu as activation function, 80 batch size and AdaMax optimizer. Best parameters were found through grid search. The denoising autoencoder uses selu as activation function and no regularizers.

\subsection{Clean Samples Classification}
\label{sec:clean_eval}

In this section, we present the performance results of MalPhase classifiers under ideal conditions, with clean samples only. Clean samples are windows of traffic containing either only benign or only malicious flows. These conditions are representative of a controlled malware analysis environment, where programs are isolated and tested individually (e.g., on a specifically-crafted VM).

\begin{figure}[t]
    \centering
    \includegraphics[width=0.9\linewidth,trim={0 3mm 0 13mm},clip]{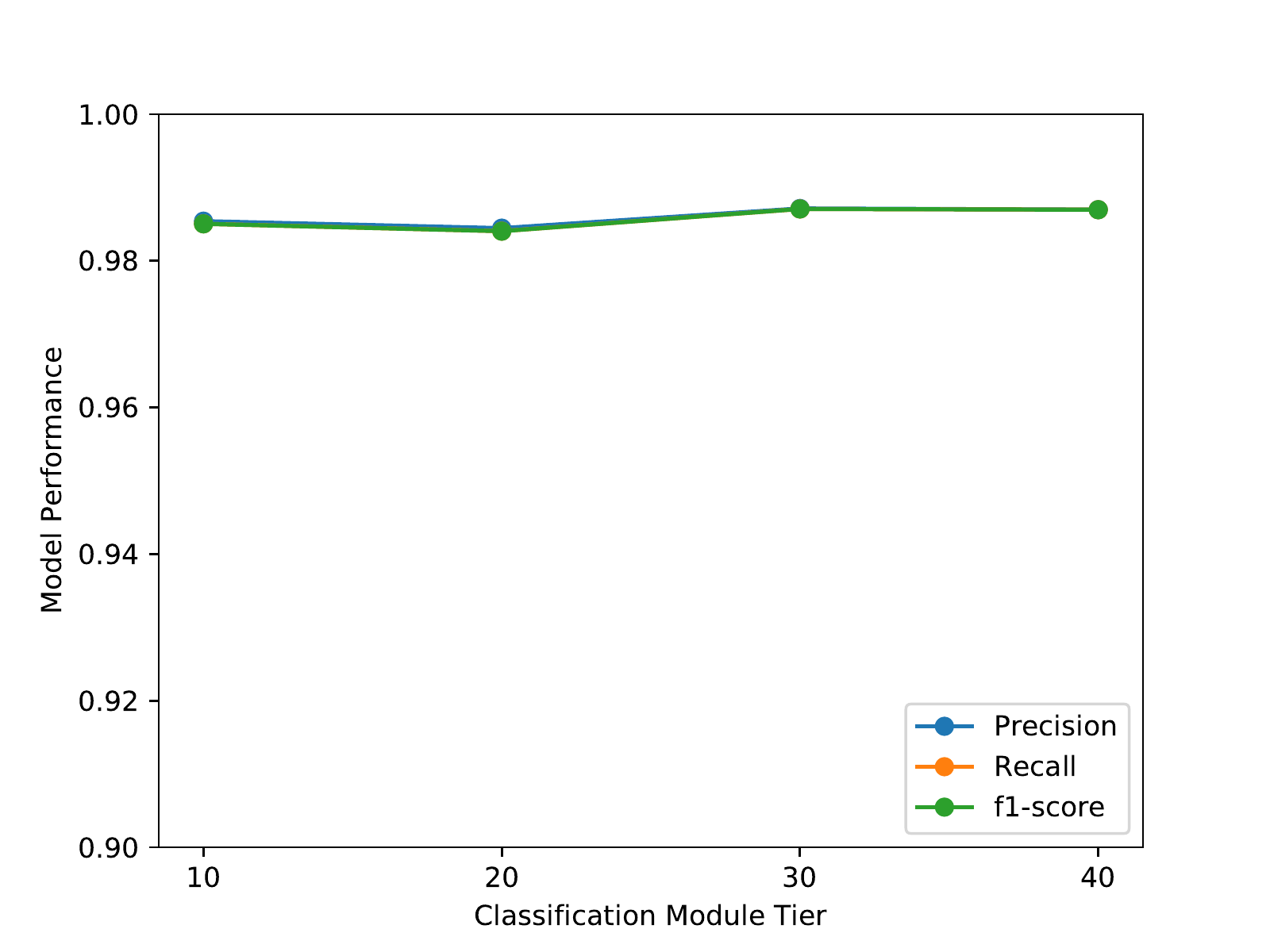}
    \caption{Precision, recall and F1-score for different tiers of the Binary Classification Modules.}
    \label{fig:cbinary}
    \vspace{-5mm}
\end{figure}

\bpar{Phase 1: Binary Classification} \\
In Figure~\ref{fig:cbinary}, we illustrate the performance of the Binary Classification Module of MalPhase on clean samples. The binary classifier exhibits very high performance in terms of the F1-score for both classes across all window sizes, with minor improvements as the window size increases.

The negligible performance difference between low and high tier classification modules is somewhat expected in this particular setting, as the evaluation is carried only on clean samples. The tiered architecture of MalPhase is designed to detect both short-term burst of malware flows, thanks to the low-tier classifiers, as well as constant, more regularly-occurring malware flows with high tier classifiers. Since samples for this evaluation contain either benign-only or malware-only flows, all malware samples can be effectively considered as bursts of malware traffic, that is quickly detected by classifiers of any tier. Moreover, higher tier classifiers are designed to exploit longer term correlations between malware flows compared to lower tier classifiers, allowing them to more precisely classify samples that exhibit similar short-term patterns.
However, when samples from different classes present very distinct behaviours --- as is the case with benign and malicious network traffic --- the models are able to easily distinguish them with only few flows, resulting in lower tier classification modules performing as well as higher tier ones. 

As we will see in Section~\ref{sec:mixed_eval} and Section~\ref{sec:tier_eval}, when considering noisy samples higher tier classifiers tend to preform better than lower tiers as the noise ratio increases.

\begin{figure}
    \centering
    \includegraphics[width=0.9\linewidth,trim={0 3mm 0 13mm},clip]{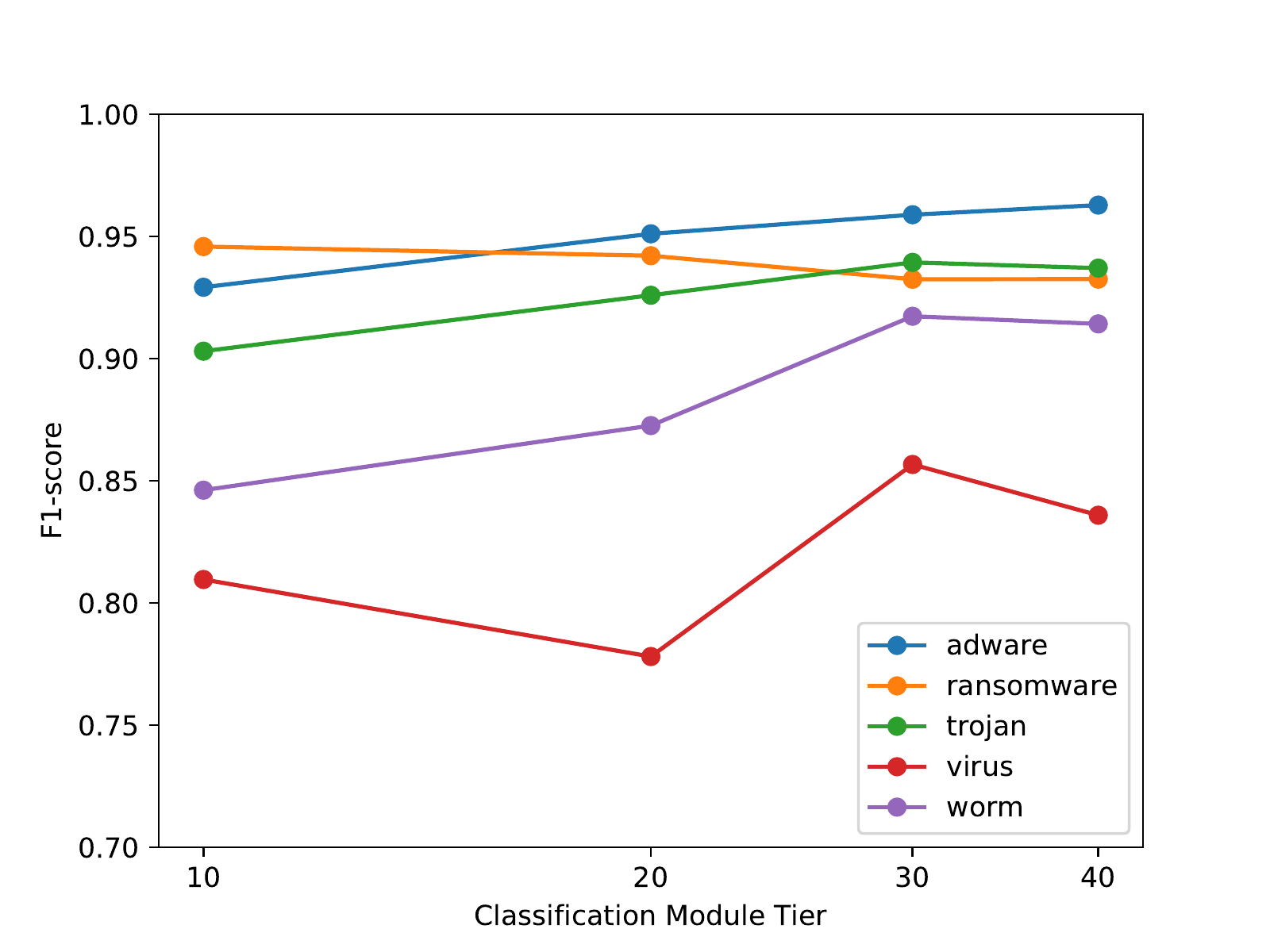}
    \caption{F1-score for different tiers of the Malware Type Classification Module.}
    \label{fig:ctype}
    \vspace{-5mm}
\end{figure}

\bpar{Phase 2: Type Classification} \\
The Type Classification Module immediately follows the binary classification, taking malicious samples as input and categorizing them as one of five malware families: adware, ransomware, trojan, virus and worm. Figure~\ref{fig:ctype} presents the performance of MalPhase's type classifier in terms of F1-score for each malware type. Overall, all classes have very good performance, with F1-score $>=80\%$ in all cases, above $90\%$ for trojan, ransomware and adware and with a weighted average F1-score of $\sim 91\%$. As could be expected, classification performance can vary considerably between different types. In particular, the classifier performs noticeably worse on identification of viruses and worms compared to the remaining types. 

This behaviour can be partially explained by the relative under-representation of these two classes in the dataset: we had three usable virus families, accounting for $\sim 5.8\%$ of the training data, and six usable worm families, accounting for $\sim 11\%$ of the training data. Another contributing factor is that assigning a type to a malware is not as clear-cut as it might appear. In recent years, malware has become increasingly complex, resulting in samples that implement behaviours from different malware types. This fact can contribute heavily to the misclassification of under-represented classes, since given a sample with cross-type behaviour, the classifier is more likely to classify it as the most represented class.

\begin{figure}
    \centering
    \includegraphics[width=0.9\linewidth,trim={0 3mm 0 13mm},clip]{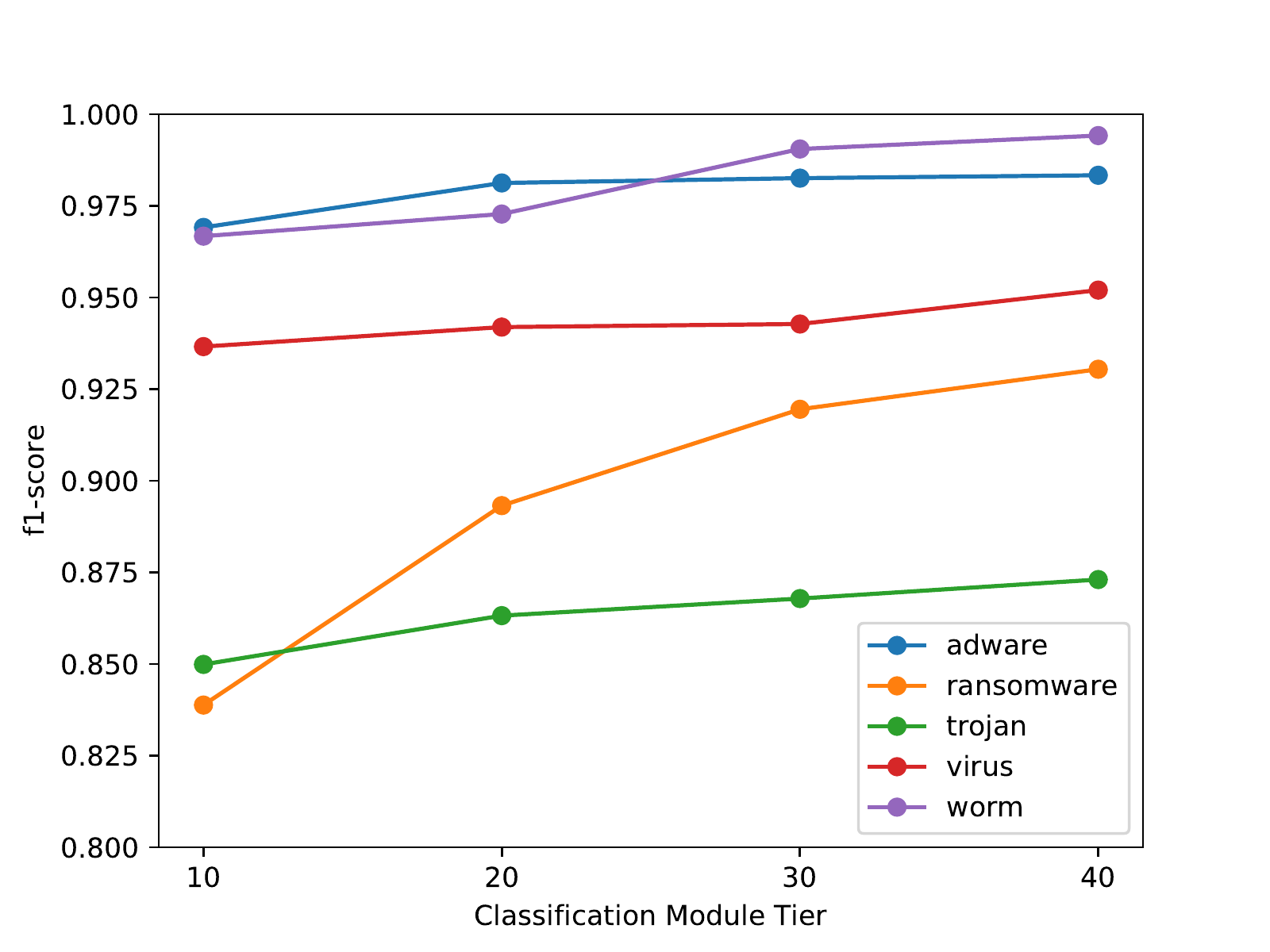}
    \caption{F1-score for different tiers of the five Malware Family Classification Modules.}
    \label{fig:cfamily}
    \vspace{-2mm}
\end{figure}

\bpar{Phase 3: Family Classification} \\
The last phase of MalPhase uses multiple type-specific classifiers to categorize the malware family. In Figure~\ref{fig:cfamily}, we show the performance for each of the five family classifiers in terms of F1-score. As we can see, most malware families are easily distinguished regardless of the classification module tier. The exception is ransomware families, with performance increasing almost by $10$ percentage points going from window size $10$ to window size $40$. This trend is indicative of the fact that on shorter flow sequences, different ransomware families exhibit behaviours that are very similar. However, as more flows are considered, the behaviours become increasingly divergent, resulting in increased performance for higher tier classification modules. 

\begin{figure}
    \centering
    \includegraphics[width=\linewidth,trim={0 2mm 0 3mm},clip]{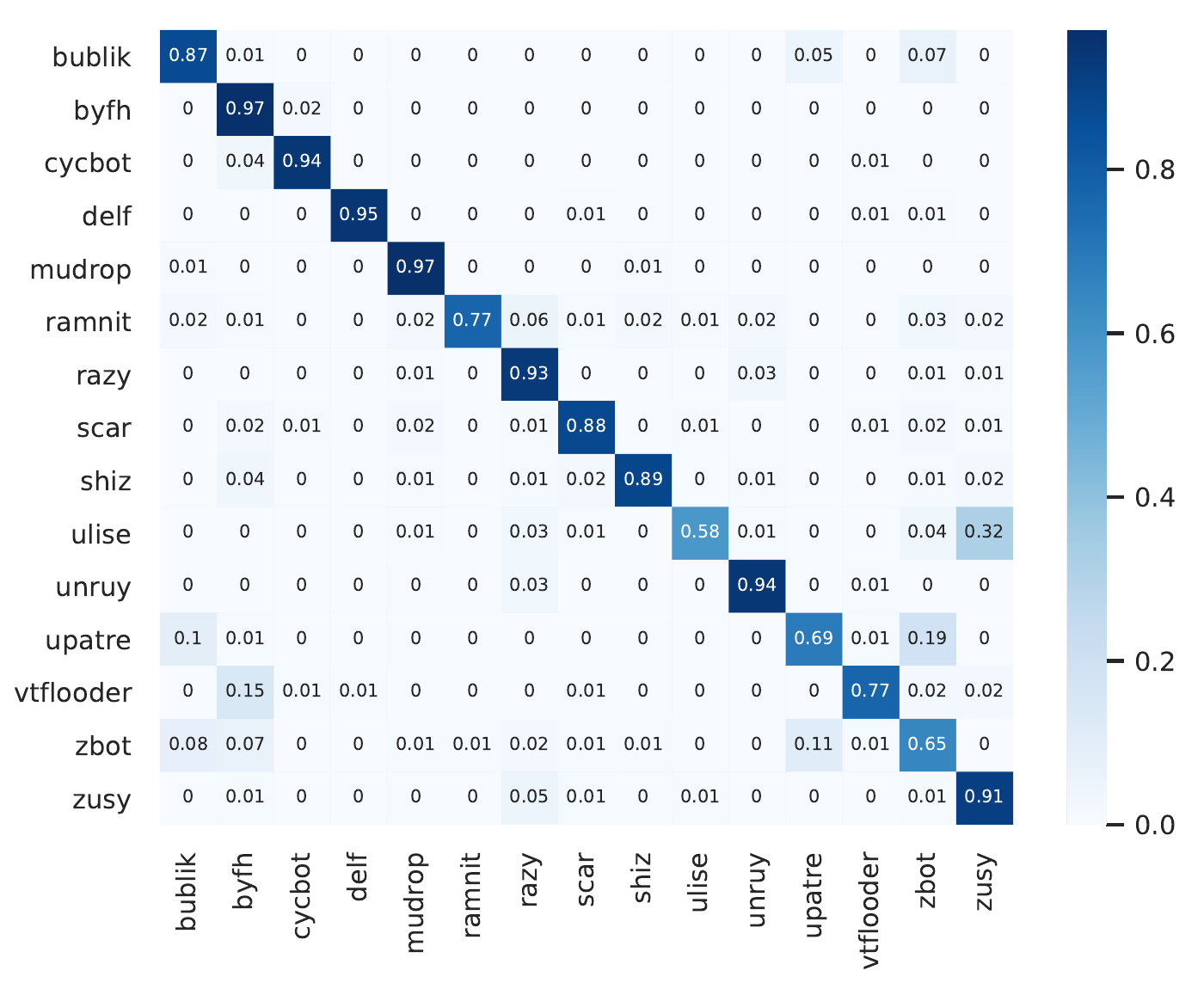}
    \caption{Confusion matrix of tier 1 family classifier for trojan family.}
    \label{fig:trojan_confusion}
    \vspace{-4mm}
\end{figure}

Interestingly, classification performance for adware families is extremely high, with F1 consistently above $97\%$ regardless of the fact that we have $9$ adware families in our dataset. On the other hand, malware families belonging to the trojan type appear to be harder to reliably classify for all classification modules, regardless of tier, with F1 hovering around $86\%$. Upon closer inspection, there are three families which are consistently misclassified by our model with recall between $60-70\%$ (see Figure~\ref{fig:trojan_confusion}): Ulise, Upatre and ZBot. 
ZBot is an alias for the famous ZeuS botnet\cite{zbot} which compromises computers using the ZeuS trojans. Upatre is a trojan downloader which, upon installation, downloads and executes additional malware on the target machine. One malware family that can be downloaded by Upatre is, in fact, ZBot itself~\cite{upatre_zeus}, which explains why the model tends to consistently misclassify samples of these two families. It's interesting to note that the number of instances in which Upatre is classified as ZBot and vice versa are rather close. This can be attributed to the fact that both classes are approximately equally represented in our dataset, resulting in similar samples classified approximately half of the time as one class and half of the time as the other.

Ulise appears to be a ``traditional'' trojan, which when downloaded on a target machine sits in the background stealing sensitive user data. Zusy, on the other hand, is an atypical piece of malware that uses a vulnerability in Office as a vector to infect a machine and steal user banking data. While having a similar goal, the two families are distinct and there doesn't appear to be any link between the two. Upon closer inspection, however, we can notice that in some cases certain AV engines mislabel Ulise samples as Zusy~\cite{ulise_zusy1,ulise_zusy2}. Since our pipeline relies on AV labeling and on the AVClass tool to assign a sample to a family (see Section~\ref{sec:dataset}), it seems likely that some samples were mislabeled due to incorrect AV/AVClass labeling. This effectively results in our model being trained on some Ulise samples that are labeled as Zusy, lowering the ability of the classifier to distinguish the two classes.

It is worth noting that during our initial dataset processing and  malware family extraction, we did not notice these correlations between different classes. While investigating  our apparent model ``misclassification'', we decided to study the correlation between these families in greater depth. This is a compelling indication of the usefulness of MalPhase, which allowed us to understand that some families (e.g., Upatre) were actually using other families as components (ZBot), resulting in very similar behaviours.

\subsection{Noisy Samples Classification}
\label{sec:mixed_eval}
This section presents the performance of MalPhase after we inject varying amounts of noise in our dataset (see Section~\ref{sec:noisy_dataset}). In this dataset, each malware sample is a noisy sample, containing a mix of interleaved malicious and benign flows. Evaluating MalPhase with noisy samples provides a closer approximation to real-life system performance, where malware traffic is interleaved with traffic from benign applications. It also gives a better understanding of the robustness of the different classification phases. We create noisy samples by randomly adding benign flows in a window of malware flows, with varying ratios of benign-to-malicious traffic (\emph{noise ratio} from here on). 

For instance, for the tier 1 classifiers that we designed to work with at least 10 malicious flows, each sample is comprised of 10 malicious flows in addition to a varying number of noise flows. For these experiments, we use noise ratios ranging from $0.2$ (i.e., benign flows are $20\%$ of malicious flows) up to $2$ for type and family classification (i.e., benign flows are $200\%$ of malicious flows), and up to $8$ for binary classification (i.e., benign flows are $800\%$ of malicious flows). Lower noise ratios are representative of bursts of malware traffic, while higher noise ratios simulate a malware that slowly and consistently performs network operations. For all experiments in this section, noise used during the training phase is sampled from a different benign dataset than noise used during testing phase. We do this, to make sure that the results were not affected by the model overfitting on the distribution of the specific noise dataset used. It is also worth noting that we did not train a specific model for every noise level, but rather we trained a single model with all noise levels to reflect a more realistic scenario.

Assessing MalPhase performance with noisy samples allows us to evaluate the effectiveness of our denoising autoencoder approach in extracting robust malware-related features to use for classification, as well as providing a much better understanding of the expected performance of MalPhase in a real-life setting.

\begin{figure}
    \centering
    \includegraphics[width=0.9\linewidth,trim={0 3mm 0 13mm},clip]{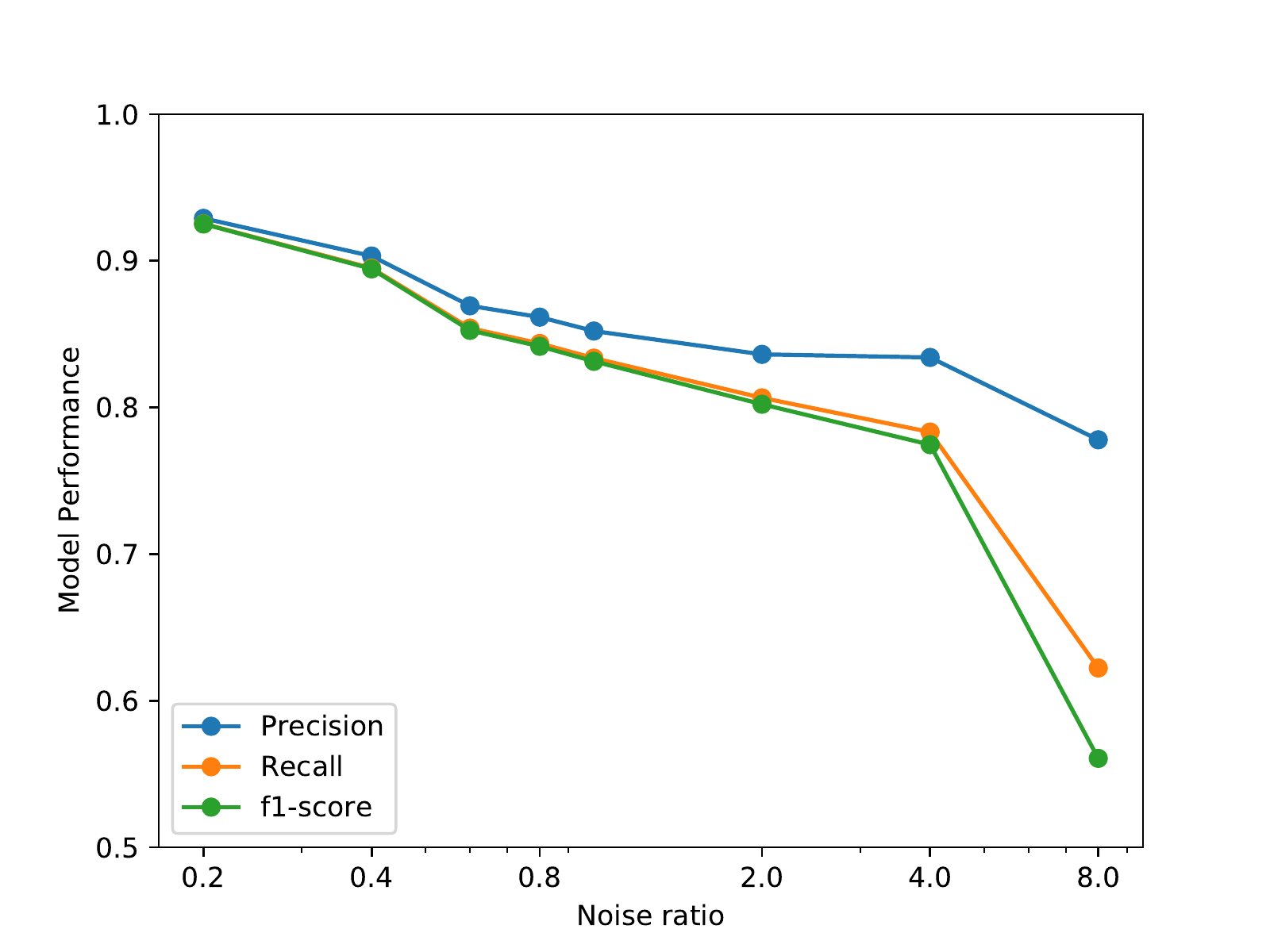}
    \caption{F1-score for the tier 3 Binary Classification Module, with varying ratio of benign-to-malicious traffic. Noise ratio plotted on a logarithmic scale.}
    \label{fig:nbinary}
    \vspace{-1mm}
\end{figure}

\bpar{Phase 1: Binary Classification} \\
In Figure~\ref{fig:nbinary}, we show the classification performance of the Binary Classification Module for tier 3, using noisy samples with varying noise ratio. The performance of the classifier decreases in a relatively stable manner between noise ratio $0.2$ and $1$, with recall and F1 going from $\sim93\%$ down to $\sim83\%$ and precision from $\sim93\%$ down to $\sim85\%$. For higher noise ratios the decrease in performance becomes less marked, with F1 decreasing to $\sim80\%$ at noise ratio $2$ and further down to $\sim77\%$ at ratio 4. This behaviour indicates most benign and malicious samples are distinct enough that even with a four-fold increase in noise ratio (from $1$ to $4$) the model can still produce very good results, with only a $6\%$ performance decrease. Beyond this point, we see that further increases in noise ratio have considerably larger impacts on performance, with F1 decreasing from $77\%$ down to $56\%$ at noise ratio $8$. This behaviour indicates that the autoencoder begins to struggle to filter out the benign noise, impacting the ability of the classifier to reliably tell apart benign and malicious samples. This general behaviour holds across all tiers, with higher tiers performing consistently better than lower tiers, as we will see later in Section~\ref{sec:tier_eval}.

\begin{figure}
    \centering
    \includegraphics[width=0.9\linewidth,trim={0 3mm 0 13mm},clip]{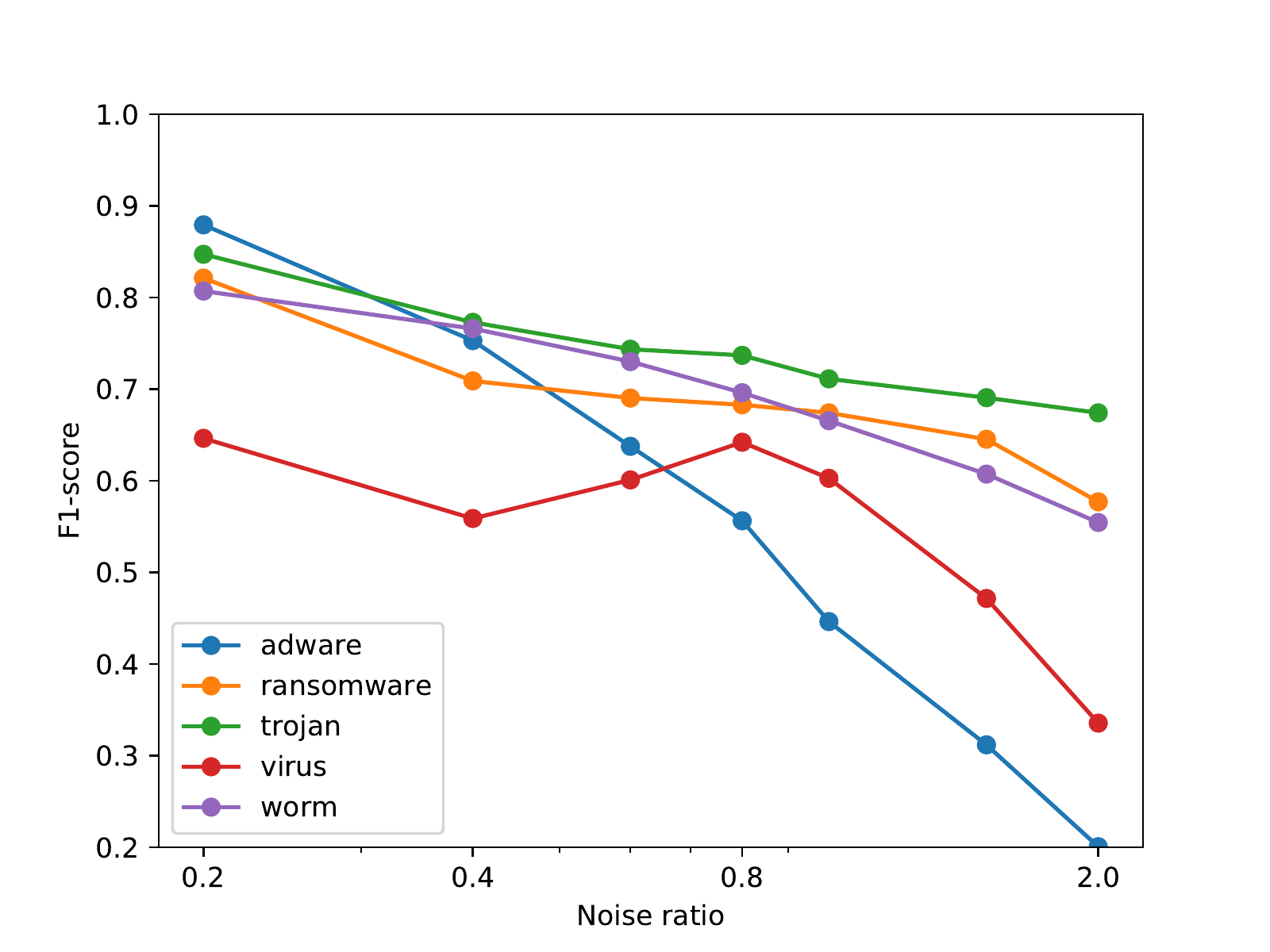}
    \caption{F1-score for the tier 3 Type Classification Module, with varying ratio of benign-to-malicious traffic. Noise ratio plotted on a logarithmic scale.}
    \label{fig:ntype}
    \vspace{-5mm}
\end{figure}

\bpar{Phase 2: Type Classification} \\
In Figure~\ref{fig:ntype}, we present the performance of the tier 3 Type Classification Module. The classifier maintains good performance on the ransomware, trojan and worm classes up to noise ratio $1$, and acceptable performance for higher noise ratios. The adware and virus classes appear to be much more sensitive to injected noise: while virus remains relatively stable for noise range $[0.2,1]$, its performance is much lower compared to that on clean samples (see Figure~\ref{fig:ctype}) and rapidly decreases for higher noise ratios. Adware performance on the other hand decreases much more sharply, losing over $30$ points between noise ratio $0.2$ and $0.8$, and decreasing even further for higher noise ratios. This result was somewhat expected for the virus class, as it is the least represented class in our dataset with just three families. On the other hand, adware is the second most represented class in the dataset with $9$ families, therefore we were not expecting the classifier to perform so poorly on it. When we look at the confusion matrix for adware, we see that it is often classified as trojan. This indicates that adware traffic is close to trojan traffic from the point of view of network flow features, and that some noise is sufficient to confuse the feature vectors enough that our model is unable to classify adware consistently.

\begin{figure}
    \centering
    \includegraphics[width=0.9\linewidth,trim={0 3mm 0 13mm},clip]{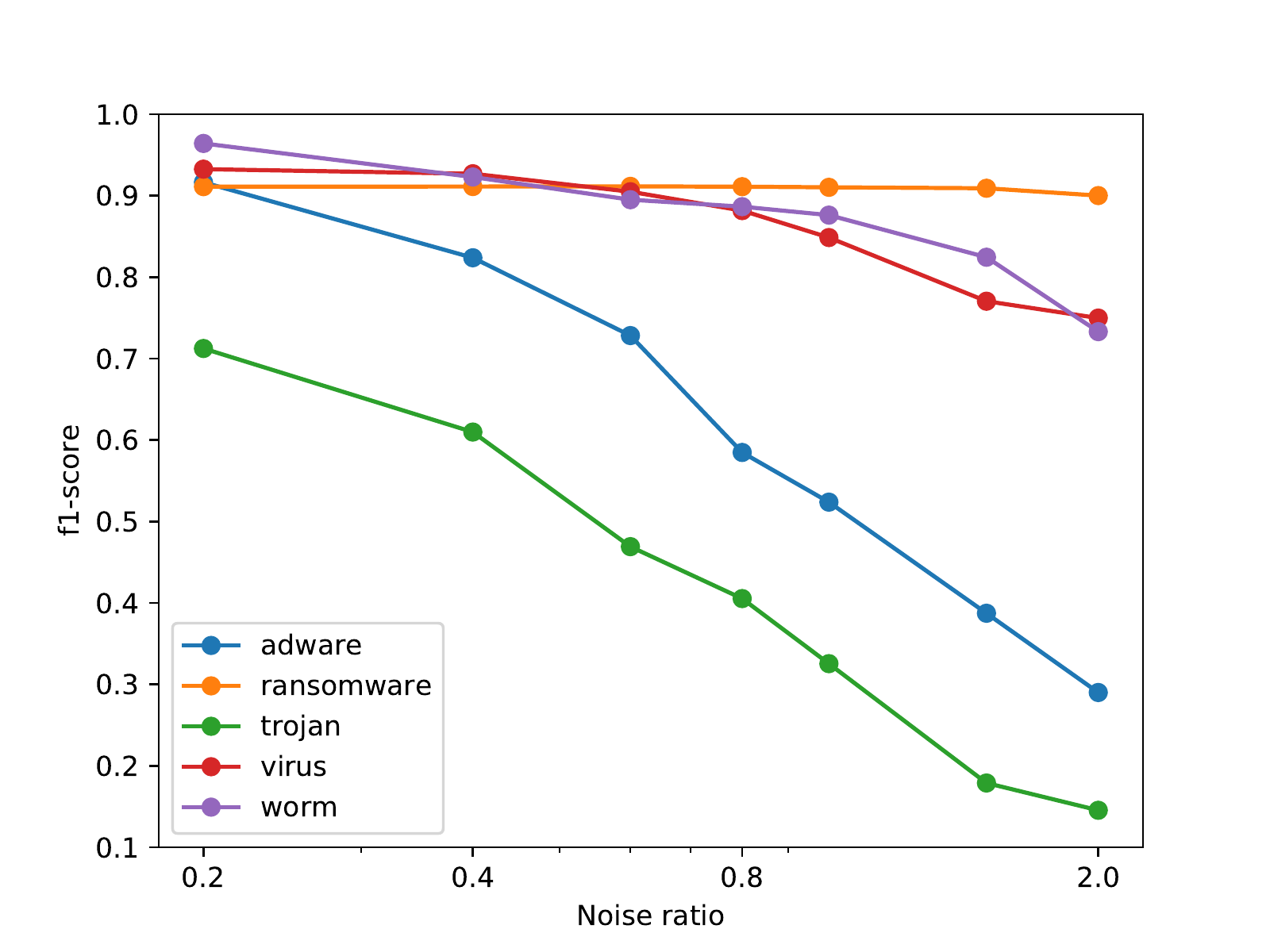}
    \caption{F1-score for the five tier 3 Family Classification Modules, with varying ratio of benign-to-malicious traffic. Noise ratio plotted on a logarithmic scale.}
    \label{fig:nfamily}
    \vspace{-4mm}
\end{figure}

\bpar{Phase 3: Family Classification} \\
In Figure~\ref{fig:nfamily}, we show the performance of the various Family Classification Modules for tier 3. The classifiers for worm, virus and ransomware are performing the best, with F1 scores around $90\%$ when the noise ratio below $1$, and consistently above $\sim75\%$ for when the noise ratio up to $2$. Trojan and adware classifiers on the other hand are much more sensitive to noise, with performance decreasing uniformly in the noise ratio interval $[0.2, 2]$. This behaviour is unsurprising, given that adware and trojan are the malware types for which we have the most families, which makes the job of the family classifier considerably more difficult. It is easy to see how, when considering ten or more target classes, any amount of noise in the input sample can severely impact the performance of the classifier. 

\subsubsection{Unseen Families}
\label{sec:unseen}
\begin{figure}
    \centering
    \includegraphics[width=0.9\linewidth,trim={0 3mm 0 13mm},clip]{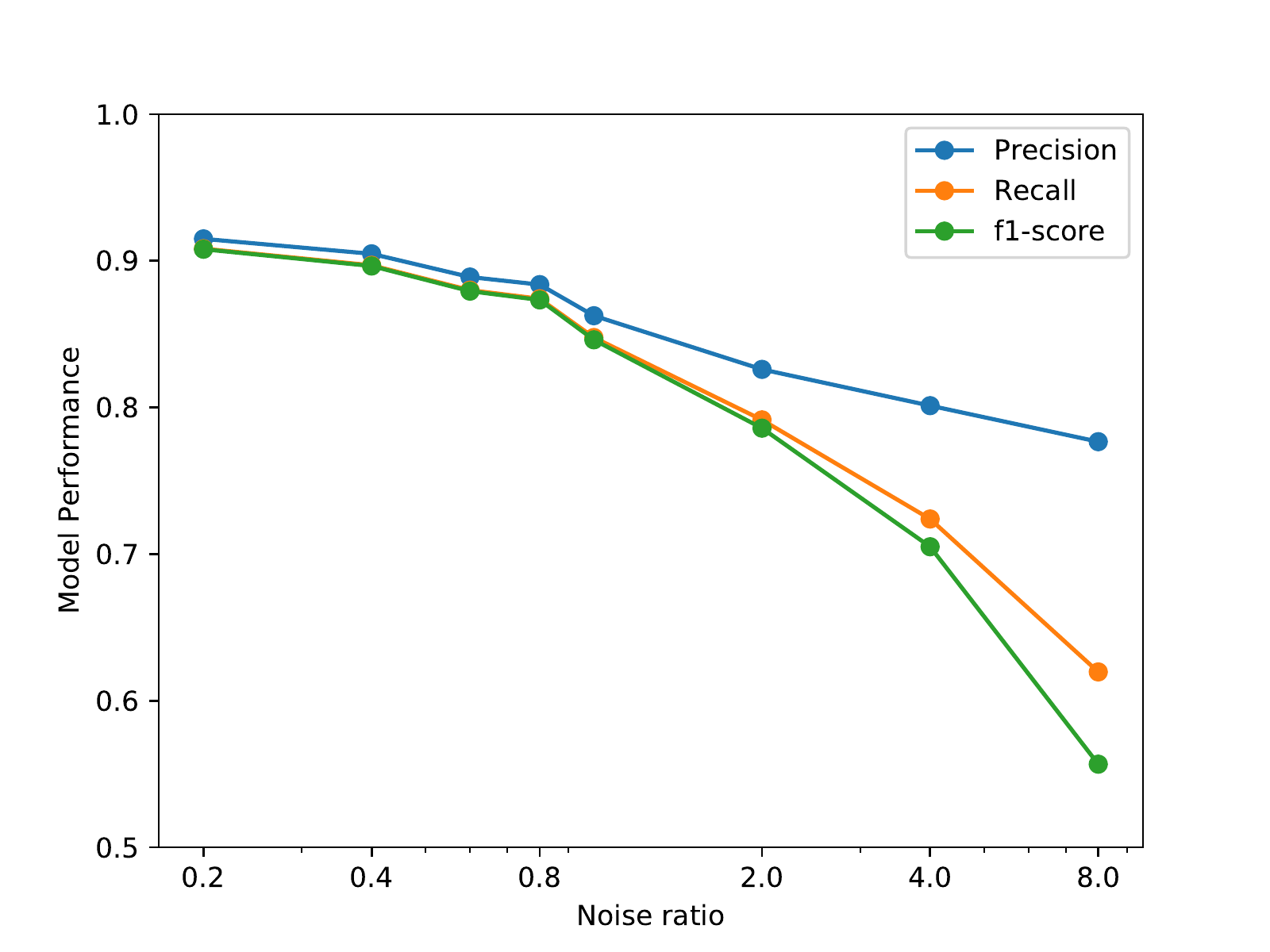}
    \caption{Precision, recall and F1-score on unseen malware families for the tier 3 Binary Classification Module, with varying ratio of benign-to-malicious traffic. Noise ratio plotted on a logarithmic scale.}
    \label{fig:unseen_binary}
    \vspace{-5mm}
\end{figure}

In this section, we examine the performance of our malware detector, the Binary Classification Module, on unseen family samples. Unseen samples are families that the classifier did not see during the training phase. Such experiment provides a better understanding of the performance the classifier would have on \emph{new, unknown malware families}. All experiments in this section are performed with noisy samples, as described in the previous section. We are therefore testing our detector in the worst case: network traffic from an unknown malware with mixed in benign traffic. For these experiments, we use the families described in Table~\ref{tab:ufamilies}. As we can see in Figure~\ref{fig:unseen_binary}, the performance of the tier 3 Binary Classification Module on unseen samples is comparable to that presented in Figure~\ref{fig:nbinary} on known samples in the noise ratio interval $[0.2,2]$, with F1-score in the range $[91\%, 78\%]$. For noise ratio above $2$, the spread in performance between known and unseen samples begins to increase, with a performance delta of $7$ points at noise ratio $4$, before decreasing again at noise $8$ where performance on known and unseen samples is comparable. This behaviour indicates that higher levels of noise have a larger impact on unseen sample classification compared to known samples. However, this penalty eventually evens out at extremely high noise levels, where the classifier performs comparably on both known and unseen samples.
Furthermore, the small difference between results on known and unseen samples suggest that improvements to performance for known noisy binary classification would most likely carry over to noisy unseen sample classification as well.

\subsection{Tier Comparison}
\label{sec:tier_eval}
\begin{figure}
    \centering
    \includegraphics[width=0.9\linewidth,trim={0 3mm 0 13mm},clip]{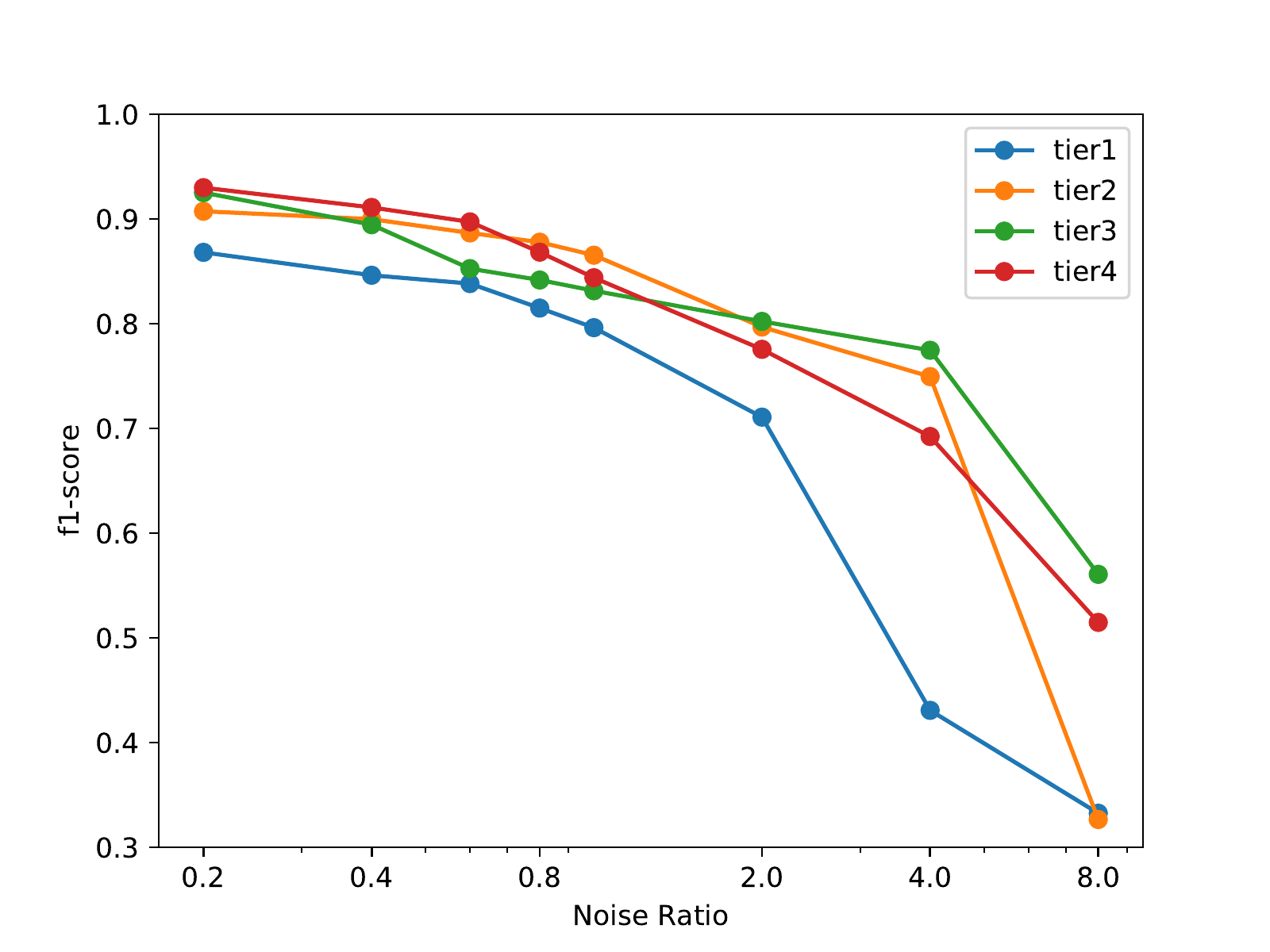}
    \caption{F1-score for Binary Classification Modules of different tier, with varying amount of noise ratio. Noise ratio plotted on a logarithmic scale.}
    \label{fig:tier_eval}
    \vspace{-2mm}
\end{figure}

In our last experiment, we assess the robustness to injection of noise for different MalPhase tiers. As discussed in Section~\ref{sec:system_architecture}, the tier system of MalPhase is designed to capture different malware behaviours: lower tiers require less malware flows to correctly classify samples compared to higher tiers, with tier 1 requiring just $10$ malware flows compared to $40$ for tier 4. This makes lower tiers suitable for the detection of quick bursts of malware traffic amidst some benign flows. Higher tiers, on the other hand, monitor a larger time horizon compared to lower tiers and are designed to be robust to higher ratios of benign-to-malicious traffic. As we can see in Figure~\ref{fig:tier_eval}, all tiers show very good performance at noise ratios between $0.2$ and $1$, with the tier 1 classifier barely falling under the $80\%$ F1-score mark. As the noise ratio increases, lower tier classifiers reach a point where performance begins to decrease sharply. Conversely, performance for higher tier classifiers decreases more gradually and reaches much higher F1-score at the maximum noise ratio. This behaviour is consistent with the rationale behind the multi-tier design of MalPhase, and shows that higher tier classifiers are more robust to noise than lower tiers, which are better suited for the detection of quick bursts of malware traffic.
\section{Novelty and Related Work}
In this section, we enumerate and analyse several different works on the subject of network-based malware detection and classification and highlight our contributions in comparison to these works.

\bpar{Scope.} One of the key contributions of our work is the use of the largest dataset of malware hashes to date for network-based malware detection, which allows us to evaluate MalPhase on a wider variety of malware families and traces when compared to previous work.
Table~\ref{gafiltad} lists the number of samples that related paper leverage in contrast to ours. Apart from Lever et al.~\cite{7958610}, the dataset used in our study is the largest to date, with several orders of magnitude more malware samples than comparable related works. It is worth noting, however, that~\cite{7958610} does not address the issue of automated malware detection and classification, but rather provides a large-scale analysis of malware from the point of view of network traffic.

\bpar{Level of Classification.} Another key contribution of our paper compared to related works is the fine-grained nature of classification in MalPhase. Indeed, many related works in this area are designed to provide only coarse grained binary detection results (i.e., benign or malicious traffic), with no type/family classification. One of the most notable recent works that falls under this category is a paper from Bartos et al.~\cite{Bartos2016}. In their work, the authors propose a new system to detect unseen malware samples based on statistical representation of features learned from ``bags of flows''. Bags of flows are an abstraction proposed by the authors which represent a set of network flows that are strictly related to each other. The authors use features extracted from proxy logs to create the bag of flows representations, and subsequently classify them in benign or malicious samples. While the goal of this work is close to our own, its scope, the level of detail of the features used for classification, as well as how this goal is achieved differ from ours. The method proposed by the authors extracts traffic features from proxy logs, which provide a richer set of information compared to flows, at the cost of increased capture complexity and scarcer availability. As previously stated, the system in~\cite{Bartos2016} is limited to detection of malware and does not perform type nor family classification. In~\cite{Ahmed2011}, Ahmed et al. propose a method to detect malicious software via analysis of network packets. While detection of malicious activity in network traffic is also part of the scope of our work, the approach proposed by the authors is aimed only at detecting malware, rather than also classifying it to type and family. Moreover,~\cite{Ahmed2011} depends on detection of executable files in packet payloads and therefore requires deep packet inspection, which makes it less practical than our flow-based approach. Disclosure~\cite{Bilge2012} performs analysis of network flows to detect botnet command and control (C\&C) servers, regardless of the protocol it employs. To maintain a low false positive rate, Disclosure augments flow data with external reputation scores. While~\cite{Bilge2012} offers a thorough evaluation on two real-world networks, the approach requires external data (reputation scores) that is not always available. The approach is specifically tailored to botnet traffic only and cannot perform classification, but only detection. Another work that is similar to ours is~\cite{Piskozub2019} from Piskozub et al. In this work, the authors propose a random forest-based classifiers based on a bag of flows representation similar to~\cite{Bartos2016}. Their classifiers is able detect malware traffic and in addition also performs malware type classification. However, the proposed approach is evaluated on a limited dataset, can only classify clean malware traffic and does not perform family classification.
\begin{table}[t]
\small
\centering
\begin{tabular}{l l r}
\toprule
\textbf{Year} & \textbf{Work} & \textbf{Malware Hashes}\\
\midrule
2007 & Tegeler~\cite{Tegeler2012a} & 188\\
2010 & Hsu~\cite{Hsu2010} & 12,629\\
2010 & Perdisci~\cite{Perdisci2019} & 25,720\\
2012 & Bilge~\cite{Bilge2012} & 37,687\\
2013 & Kheir~\cite{Kheir2013} & 2,143\\
2013 & Rafique~\cite{Rafique2013a} & 15,850\\
2014 & Invernizzi~\cite{Invernizzi2014} & 43,380\\
2014 & Mohaisen~\cite{Mohaisen2014} & 2,699\\
2016 & Bartos~\cite{Bartos2016} & 7,000\\
2017 & Lever~\cite{7958610} & 26,800,000\\
2018 & Deng~\cite{Deng2018} & 999\\
\midrule 
2021 & This paper & 13,920,730 \\
\bottomrule
\end{tabular}
\caption{Malware Sample Size of Comparable Works}
\label{gafiltad}
\vspace{-5mm}
\end{table}

\bpar{Scope of Classification.} One important aspect that distinguishes our proposal from previous works is the broad scope of classification. While there are related works that perform fine-grained malware family classification similarly to MalPhase, these approaches are generally applicable only to specific malware families (i.e., botnet). In contrast, MalPhase family classification is not limited to a specific malware type and can generalize. In~\cite{8844609}, Mar{\'i}n et al. assess the applicability of deep learning models to malware detection and classification using raw bytestream-based data. In this work, the authors focus on a small subset of 3 botnet families and show that their model is able to detect and classify them with high performance. The use of such a small number of families, as well as the limit of working only on botnet traffic heavily limits the applicability of their classifiers compared to ours. Gezer et al.~\cite{Gezer2019} propose a flow-based machine learning approach to detect the banking trojan TrickBot. The approach proposed by the authors heavily differs from ours in term of features, architecture, as well as in the reduced scope of the paper. Gu et al. BotHunter~\cite{Gu2007} is a well-known system that performs packet-level analysis to enable bot identification, based on a state-based infection sequence model of botnet behaviour. Similarly, BotFinder~\cite{Tegeler2012a} is an exploratory research system that uses high level traffic features to detect the presence of botnet traffic from 6 specific families. Botminer~\cite{Gu2008} is an approach to detect botnet based on unsupervised clustering and cross-cluster correlation that is able to detect the presence of botnet malware in network traffic. While these works have a similar goal to ours, they only provides coarse grained results (detection only, no classification), are limited to a small number of families and are designed to work only on botnet traffic. Botection~\cite{Alahmadi2020} is a machine learning-based system that uses network flows to detect and classify botnet families, building on earlier work on behavioural analysis on network flows by the same authors~\cite{Alahmadi2018a}. Botection relies on Markov Chains to represent the bot network comunication sequence and preform detection and classification among 17 botnet families. While Botection shows the ability to generalize to a larger set of families compared to most previous works, its scope is much more narrow compared to MalPhase. In fact, the classifier in~\cite{Alahmadi2020} is designed to use specific state transitions sequences for classification. These state transition features that are handcrafted and designed to work for botnets only, while MalPhase detection and classification capabilities are general and not restricted to a single malware type.
 
\bpar{Robustness.} Finally, a key contribution of our work is the robustness of MalPhase classification. As a result of the use of denoising autoencoders, MalPhase classification is resilient to noise injected in the malicious malware trace. To the best of our knowledge, MalPhase is the first system that combines fine-grained classification for a broad range of malware types and resilience to noise. A related work in this aspect is Botection~\cite{Alahmadi2020}, which the authors show is able to detect botnet malware even when some noise is injected in the malware trace. However, MalPhase exhibits comparable or better performance in noisy binary detection, while also working on a broader set of families (more than double) and types compared to only botnets for~\cite{Alahmadi2020}. Moreover, MalPhase can also perform type and family classification on noisy traffic, albeit with decreasing performance as the noise ratio increases. 

\bpar{Other.} There are many other works that apply machine learning to malware traffic to achieve goals other than detection and classification. Shibahara et al.~\cite{7841778} apply deep learning techniques to automate the collection of network traces in an efficient and effective manner. In~\cite{Zhu2016} Zhu et al. describe a natural language processing-based method to learn malware behaviour from academic literature and automatically extract meaningful features for malware detection. Perdisci et al.~\cite{Perdisci2019} propose an unsupervised clustering method to group similar malware families and automatically generate signatures for detection. 
Similarly, Firma~\cite{Rafique2013a} clusters traffic into particular families and then generates IDS signatures for those families. While these work belongs to the same general area of ML-based malware approaches, these systems do not perform detection and classification and require a much wider range of features.
Some previous works in the literature use entropy similarly to us to enrich network flow data. BotHunter~\cite{Gu2007} uses entropy in its packet-level analysis as an indicator of maliciousness of the network traffic. Do et al.~\cite{9053419} describe a machine learning-based technique to classify a variety of network attacks based entropy of network flows. Entropy of binaries has also been reliably used as an indicator of maliciousness~\cite{4140989,Shafiq2008}, although recent research has shown that this assumption may be flawed~\cite{Mantovani2020} regarding packers. This is similar to network forensics, wherein it can be assumed that entropy is an indicator of encryption and therefore maliciousness~\cite{degaspari_encod}. These works are only loosely related to our proposal, as they give the intuition that integrating measures of entropy as part of the features can potentially improve detection and classification results.

\section{Limitations}
\label{sec:discussion}
\bpar{Malware Taxonomy.}
We found the lack of clearly defined malware taxonomies to be one of the obstacles in creating a robust malware dataset. A number of papers define ways to categorise malware. There is a general consensus that different strains or mutations of malware are classified into a family, while type (or class) is a more general category based on malware modus operandi. To add complexity to the process, sometimes one of those three terms (family, type and class) is confused for the other one. For instance, Rhode et al.~\cite{Rhode2018} use the term `family' when they talk about types; AVClass~\cite{10.1007/978-3-319-45719-2_11} uses the term `class' for its output, which is a malware family. The more high-level the categorisation, the more liberty and flexibility is assumed by paper authors.
Some of them adopt similar malware types to ours, while combining malware with attacks in their taxonomy~\cite{sperotto2010}, or provide a simple taxonomy that divides malware into simple and self-reproducing~\cite{filiol}. Gr{\'{e}}gio et al.~\cite{Gregio2014} assume yet another type taxonomy, treating viruses, worms and trojans as a single type. Additionally, AV vendors provide their own fixed set of malware types, which varies depending on the specific company. The most comparable type taxonomy to ours is the one presented by Cisco~\cite{ciscoMalware}, with the addition of adware to the list of ransomware, viruses, worms and trojans. A notable and deliberate omission on our part, is the botnet type. Its main characteristic is the way in which it interacts with its owner, from whom it takes new commands. The term 'botnet' does not define the way of operation of malware or its behaviour (i.e., there are worm botnets, trojan botnets or virus botnets), which was the main criterion in our selected type taxonomy. Having said this, botnets are still a valid way of defining certain malicious programs, however we feel they belong to a separate taxonomy.

\bpar{Network Flows.}
MalPhase uses bidirectional flows, which contain all standard features such as protocol, source and destination IP addresses and ports, packet sizes and counts. The flows are augmented with additional features: round-trip time of packets, and packet payload entropy. While the calculation of Shannon entropy requires access to packet payloads, it is still privacy-preserving. Due to those additional features our method does not operate on default flows (e.g., NetFlows) with the same classification performance. Additionally, we do not use IPv6 traffic in our datasets and evaluation, however including them would not require significant changes to our system.

\bpar{Evasion.}
Many malware variants today make use of evasion techniques in order to avoid detection. One such technique consists of injecting benign traffic to obfuscate the malicious traffic generated by the malware itself~\cite{Deng2018,Tegeler2012a,10.5555/2028067.2028096}. As highlighted in Section~\ref{sec:mixed_eval}, MalPhase shows good robustness to injection of benign traffic. However, injection of high amounts of benign traffic decreases the performance of the classifiers, with a more pronounced impact on type and family classification. 
Other host-based malware evasion techniques such as mimicry of benign processes can be used, though they do not always evade network forensic analysis ~\cite{degaspari2020}.

\bpar{Noise Injection.}
Noise injection provides a useful indication of the robustness of the classification approach and its applicability in a real world setting, but it also has some limitations. Whilst the noise is sampled from real world benign traffic, randomly injecting it in a window of malicious flows does not necessarily preserve the temporal properties that the benign data would have in the real world. There is no guarantee that the benign traffic in our dataset captures the true distribution of benign traffic in general. We believe noise injection is an acceptable compromise to assess the resilience of MalPhase.

\bpar{Unseen Type Classification}
While our evaluation shows that MalPhase is able to detect unseen malware samples with high performance, we were unable to replicate the same result for type classification. We believe that this limitation stems from an insufficient amount of families left after dataset processing (see Table~\ref{finalFamilies}). It is possible that the amount of available families is insufficient for the model to learn a type representation that can generalize to unseen samples. However, a higher number of families per type would also likely impact the performance of the classifier in the family classification task.

\bpar{Sandboxes.} As all of the used malware network traffic data is collected from sandboxes, our dataset is biased towards malware that does not employ anti-VM techniques~\cite{Miramirkhani2017,Yokoyama2016}. Since network traffic was captured in the sandbox only for a short length of time, malware that deploys delayed execution may not leave any forensic artefacts, and therefore be excluded from our dataset.

\section{Conclusions}
\label{conclusions}
We present MalPhase, a multi-phase system for the detection and classification of malware network traffic, and evaluate it on one of the largest datasets of malware flows to date. MalPhase is capable of detecting malware from aggregated network flow data, as well as classifying it into  specific malware type and family. We show that MalPhase performs comparably or better than current state-of-the-art approaches on clean malware sample detection ($>98\%$ F1), while also broadening the scope of classification to malware type ($>93\%$ F1) and family ($>91\%$ aggregated F1). Furthermore, we evaluate and demonstrate the robustness of MalPhase classification to the injection of benign flows interleaved with malware network traffic. Finally, we show that MalPhase is able to detect unseen samples with performance comparable to that of known samples, even in the presence of noise.

\section*{Acknowledgements}
The authors would like to thank VirusTotal and Google Cloud Security for providing much of the source data.
\balance

\bibliographystyle{IEEEtranS}
\bibliography{bibmalphase}
\end{document}